\documentclass{article}
\usepackage{fullpage}
\usepackage[utf8]{inputenc}
\usepackage[T1]{fontenc}    
\usepackage{hyperref}       
\usepackage{booktabs}       
\usepackage{amsfonts}       
\usepackage{nicefrac}       
\usepackage{enumitem}
\usepackage{amsmath}
\usepackage{amssymb}
\usepackage{amsthm}
\usepackage{xcolor}
\usepackage{natbib}
\usepackage{graphicx}
\usepackage{subcaption}
\usepackage{hanging}
\usepackage{authblk}

\newcommand{\diff}{\Delta}

\newcommand{\ICML}{ICML}
\newcommand{\ICMLyear}{ICML 2020}
\newcommand{\papidx}{i}
\newcommand{\revidx}{j}
\newcommand{\strategy}{S}
\newcommand{\positive}{\strategy^{+}}
\newcommand{\negative}{\strategy^{-}}

\newcommand{\numpap}{m}

\newcommand{\specexppap}{\mathcal{M}^{\ast}}
\newcommand{\exppap}{\mathcal{M}}
\newcommand{\tempset}{\mathcal{T}}
\newcommand{\score}{\theta}

\newcommand{\papload}{\lambda}

\newcommand{\posgroup}{\exppap_{+}}
\newcommand{\neggroup}{\exppap_{-}}
\newcommand{\specposgroup}{\specexppap_{+}}
\newcommand{\specneggroup}{\specexppap_{-}}

\newcommand{\frst}{{\sc initiating}}
\newcommand{\scnd}{{\sc balancing}}

{
\theoremstyle{definition}

\newtheorem*{remark*}{Remark}

}

\title{A Large Scale Randomized Controlled Trial on Herding in Peer-Review Discussions}

\author[${}^\spadesuit$]{Ivan Stelmakh}
\author[${}^\spadesuit$]{Charvi Rastogi}
\author[${}^\spadesuit$]{Nihar B. Shah}
\author[${}^\spadesuit$]{Aarti Singh}
\author[${}^{\heartsuit \diamondsuit}$]{Hal Daum{\'e} III}

\affil[${}^\spadesuit$]{School of Computer Science, Carnegie Mellon University}
\affil[${}^{\heartsuit}$]{University of Maryland, College Park}
\affil[${}^{\diamondsuit}$]{Microsoft Research, New York}

\date{\vspace{-13pt} \texttt{\{stiv,crastogi,nihars,aarti\}@cs.cmu.edu, hal@umiacs.umd.edu} }

\begin{document}

\maketitle

\begin{abstract}
Peer review is the backbone of academia and humans constitute a cornerstone of this process, being responsible for reviewing papers and making the final acceptance/rejection decisions. Given that human decision making is known to be susceptible to various cognitive biases, it is important to understand which (if any) biases are present in the peer-review process and design the pipeline such that the impact of these biases is minimized. In this work, we focus on the dynamics of between-reviewers discussions and investigate the presence of herding behaviour therein. In that, we aim to understand whether reviewers and more senior decision makers get disproportionately influenced by the first argument presented in the discussion when (in case of reviewers) they form an independent opinion about the paper before discussing it with others. Specifically, in conjunction with the review process of \ICMLyear{} --- a large, top tier machine learning conference --- we design and execute a randomized controlled trial with the goal of testing for the conditional causal effect of the discussion initiator's opinion on the outcome of a paper.  
\end{abstract}

\section{Introduction}
\label{section:intro}

While carefully aggregated independent opinions of a large number of people can often result in superior accuracy~\citep{galton07vox, surowiecki05wisdom}, decisions stemming from a \emph{a group discussion} are known to be susceptible to various biases related to social influence~\citep{asch51conformity, baron96conformity, lorenz10nowisom, janis1982groupthink, cialdini04influence}. For example, a seminal experiment of~\citet{asch51conformity} revealed that when a strikingly incorrect judgement is made by several group members, other agents may agree with it even if they clearly understand its inaccuracy. Manifestations of social influence have been documented in political elections~\citep{bond12mobilization}, individual and institutional investments~\citep{nofsinger99herding}, and academia~\citep{resnik20groupthink}.  

This work considers a specific manifestation of social influence that results in \emph{``herding behaviour''} --- an effect when agents are doing what others are doing rather than choosing the course of actions based on the information available to them~\citep{Banerjee92asimple} --- in group discussion. A long line of work on human decision making establishes the presence of various biases related to the first piece of information received by an individual~\citep{tversky74heuristics, strack97anchoring, mussweiler01achoring}. In particular, these works show that an initial signal received by a decision maker (even when being clearly unrelated to the underlying task) often has a disproportionately strong influence on the final decision. Projecting this observation on the group discussion setting, we hypothesize that the first argument made in the discussion may  exert an undue influence on opinions of subsequent contributors, thereby leading to a herding behaviour.

Past literature on group decision making~\citep{McGuire87group, dubrovsky91status, weisband92advocacy} indeed suggests that the herding behaviour (titled ``first advocacy'' effect in these works) may be present in group discussions.\footnote{We discuss these past works in more detail in Section~\ref{section:discussion}.} With this motivation, in this work we concentrate on a specific incarnation of the discussion process and analyze the presence of the herding effect in the the discussion stage of conference peer review. In peer review, discussion takes place after reviewers submit their initial reviews and authors submit their responses to these reviews. The purpose of the discussion stage is to allow reviewers and area chairs (equivalents of associate editors in journal peer review) to exchange their opinions about the papers and correct each other's misconceptions. Overall, reviewers are supposed to reach a consensus on the paper or boil their disagreement down to concrete arguments that can later be evaluated by area chairs and program chairs (equivalents of the editor in chief in journal peer review).

Given that discussion plays a central role in conference peer review and that many aspects of conference peer review such as bidding~\citep{fiez2019super,meir2020market}, paper-reviewer matching~\citep{stelmakh2018pr4a,garg2010papers, kobren19localfairness, jecmen2020manipulation}, rebuttals~\citep{gao19rebuttal}, review ratings~\citep{tomkins17wsdm, kang18peerread, noothigattu2018choosing, wang2018your} and rankings~\citep{stelmakh2020catchme,xu2018strategyproof}, as well as review text~\citep{hua2019argument,manzoor2020uncovering}  are studied by past literature, it is perhaps surprising that the discussion between reviewers remains largely unstudied. The work of~\citet{gao19rebuttal} performs an analysis of observational data on post-rebuttal review update and concludes that reviewers tend to converge to the mean of scores given in initial independent reviews. This finding is supported by a careful randomized controlled study by~\citet{teplitskiy19influence} which indicates a strong impact of social influence on experts' evaluations. However, these past works do not investigate the dynamics of the between-reviewer discussion that we consider in this work. 

The most relevant past work on peer review is a study by~\citet{hofer00care} which evaluates the impact of discussion in peer review of medical care quality on the accuracy of final decisions. The experiment by~\citet{hofer00care} reveals that while discussion between a pair of reviewers leads to higher agreement within the pair, it does not improve the agreement between reviewers from different discussion pairs, suggesting that the inflation of the post-discussion agreement does not indicate an improvement of the quality of resulting decisions. This finding hints that reviewers within the pair reach a consensus not because they identify the right answer, but due to some other effects, and in the present study we aim at investigating the presence of the herding behaviour --- a potential cause of the effect observed by~\citet{hofer00care}. 

In addition to being a potential cause of erroneous consensus between reviewers, the herding behaviour, if present, can contribute to the overall unfairness of the review process. In the current review system, it is the job of the area chairs to ensure that discussions do take place, and different area chairs use different strategies to initiate the discussion. Some of them may call upon the reviewer whose opinion is the most extreme, others may request the most positive (negative) reviewer to start. Another option for an area chair it to initiate the discussion themselves or to choose an initiator based on their seniority or expertise.  In the presence of herding, the uncertainty in the choice of the strategy may impact the outcome of a paper (which becomes dependent on the essentially random choice of the strategy by the area chair), thereby increasing the undesirable randomness of the process. 

With the above motivation, in this work we aim at testing the presence of the herding effect in conference peer-review discussions. Specifically, in conjunction with the review process of the International Conference on Machine Learning (\ICMLyear{}) --- a flagship machine learning conference that receives thousands of paper submissions and manages a pool of thousands of reviewers --- we study the following research question:
 
\begin{center}
\begin{minipage}{0.9\linewidth}
  \begin{hangparas}{8.8em}{1}
    \textit{Research Question:} Given a set of reviewers who participate in the discussion of a paper, does the final acceptance decision for the paper {causally} depend on the choice of the discussion initiator?
  \end{hangparas}
\end{minipage}
\end{center}

Let us now clarify some subtle points in the formulation of the research question. First, by saying that we are interested in a causal relationship, we underscore that our goal is to exclude the impact of the various confounding factors that are present in the observational data on peer-review discussions. For example, the fact that some reviewer decides to begin the discussion may imply that this reviewer is the most energetic and hence would dominate the discussion even if it was initiated by someone else, thereby introducing spurious correlations in data. We design the experiment to remove such spurious correlations and \emph{identify the causal impact of the order in which reviewers join the discussion.}

Second, by saying that we consider the effect conditioned on reviewers who participate in the discussion, we specifically address the confounding factor that not all assigned reviewers may participate in the discussion of a paper. Hence, any intervention may impact the outcome of the paper by altering the set of reviewers involved in the discussion, thereby also leading to spurious correlations. As we explain in the sequel (Section~\ref{sec:design}), we carefully design an intervention that allows us to  account for this undesirable effect.

Finally, despite past work finding herding in human discussions, these findings may not necessarily apply to peer review because of the nature of the task performed by reviewers and presence of other biases. Indeed, the reviewing task is analytical and requires rational thinking, thereby potentially reducing the reliance on heuristics responsible for cognitive biases~\citep{stanovich99rational, kahneman02representativeness}. On the other hand, the discussion takes place after the initial reviews are submitted and hence reviewers may be anchored to opinions they have already formed~\citep{tversky74heuristics}, which could also reduce the strength of the potential effect.

\section{Experiment Design}
\label{sec:design}

In this section, we describe the design of the experiment we executed in \ICMLyear{} to test for the herding effect in conference peer review. Before we delve into details, let us briefly describe the organization of the \ICMLyear{} peer-review process, which follows the conventions adopted by most top machine learning and artificial intelligence conferences. The conference peer-review pipeline comprises four main components: (i) initial reviewing, (ii) author rebuttal, (iii) reviews update and discussion, (iv) final decision making. Upon the release of initial reviews, authors of papers have several days (12 days in the case of \ICMLyear{}) to write a response to reviewers, followed by the discussion stage. During the discussion, reviewers and area chairs have access to the author feedback and are able to communicate with each other (but not with authors) via a special interface. For the papers assigned to them, each reviewer is expected to carefully read the author rebuttal as well as the reviews written by the other reviewers, and try to reach a consensus with other reviewers on the final recommendation for the paper or identify the key points that require an evaluation of the area and program chairs. Area chairs are responsible for overseeing and managing discussions; specifically, they have to ensure that a discussion between the reviewers (as well as the area chair) does take place for papers that require the discussion (i.e., for which the final decision is not yet clear) and they are free to use any strategy to initiate it. After the discussion is over, the area and program chairs analyze the author responses, reviews and discussions to make the final decisions.

\subsection{Preliminaries}
\label{section:preliminaries}

To establish a \emph{causal} relationship between the opinion of a discussion initiator and the outcome of a paper, the experiment we design in this work follows an  A/B testing pipeline in which a set of papers submitted for review is split into two groups that receive different treatments, where the treatments are designed such that the difference in some observable outcome of papers across groups is indicative of the presence of herding.

Intuitively, we expect herding (if present) to move the outcome of a discussion towards the opinion of the discussion initiator. Recall that discussions begin \emph{after} the initial reviews are submitted, thereby enabling us to infer the initial opinions of reviewers about the papers. Thus, a na\"ive idea to conduct such a test is to assign the reviewer with the most positive (respectively, negative) opinion about the paper to initiate the discussion for each paper from group A (respectively, B). In this way, under the presence of herding we expect papers from group A to receive more positive final evaluations from reviewers and enjoy higher acceptance rate than papers from group B, thereby allowing us to detect the effect. However, as compared to the standard A/B testing settings, the conference peer review process features important idiosyncrasies that complicate the design of the treatments:
\begin{itemize}[itemsep=5pt, leftmargin=15pt, topsep=5pt]
    \item First, in contrast to some conventional experiments on the herding behaviour~\citep{McGuire87group, dubrovsky91status, weisband92advocacy} where subjects are specifically recruited to participate in the discussion and hence the participation rate is high, in peer review some reviewers may choose to ignore the discussion. In fact, the analysis of the review process of another leading machine learning conference NeurIPS 2016~\citep{shah2017design} revealed that only  30\% of 13,674 paper-reviewer pairs had a message posted by the reviewer in the associated discussion, showing that the set of discussion participants is generally a strict (and somewhat small) subset of reviewers assigned to the paper. 
    
    \item Second, even if a reviewer is willing to participate in the discussion, they may refrain from initiating the discussion when requested to do so by the area or program chairs. Moreover, while we can encourage some reviewers to start the discussion, we cannot prevent other reviewers from doing so if they wish. Overall, we do not have complete control over who initiates the discussion.
    
\end{itemize}

\noindent The first idiosyncrasy implies that any treatment we apply to a paper may affect the outcome of the paper in two ways: first, by inducing a specific (negative or positive) shift in reviewers' evaluations caused by herding (the intended way described above)  and second, by changing the population of reviewers participating in the discussion (an unintended way orthogonal to herding that can obscure the findings). To understand the second mechanism, note that under the na\"ive treatments introduced above, the participation rate of the the most positive (respectively, negative) reviewers may be higher for papers from group A (respectively, B) because our treatments specifically encourage these reviewers to participate in the discussion of papers from the corresponding groups. The difference in the participation rates may affect the outcome of a paper even in the absence of herding, thereby leading to an increased probability of the false alarm. 

The second idiosyncrasy further complicates the design by limiting our control over the choice of the discussion initiator; even if the we request some reviewer to initiate the discussion, there is no guarantee that the reviewer will comply. To illustrate the potential consequences, let us consider a concocted scenario in which the most positive and the most negative reviewers are reluctant to initiate the discussion and where all the discussions are initiated by reviewers whose scores are not at the extremes. In this example, the na\"ive treatments we introduced above do not induce any difference between the conditions and, hence, the collected data will not allow to identify the herding effect even if it is present.

Taking these challenges into account, we now proceed to describe the design of the intervention that we implement in the \ICMLyear{} peer-review process.

\subsection{Intervention Design}
\label{sec:instrument}

\newcommand{\posrevconv}{R^{+}}
\newcommand{\negrevconv}{R^{-}}

Let us refer to the set of papers that are involved in the experiment as ``participating papers'' (the selection criteria is explained in Section~\ref{sec:exppr}). Recall that the idea of our intervention is to split the set of participating papers into two subgroups (A and B) and apply different treatments to each of these groups. In this paper, the treatments we apply to the groups are specific discussion-management strategies, so in what follows we use the latter term.

To clearly formulate the requirements on the strategies applied to the different groups of papers, let us consider a paper and two scenarios of its discussion that correspond to allocation of the paper to group A or group B. To account for the idiosyncrasies formulated in the previous section, the strategies we design should ensure that:
\begin{enumerate}[topsep=2pt, label=(\roman*)]
    \item The set of reviewers who participate in the discussion of the paper is the same in both scenarios.
    \item The most positive (respectively, negative) reviewer is more likely to initiate the discussion when the paper is allocated to group A (respectively, B) than other reviewers.
\end{enumerate}
In other words, the strategies should ensure that set of reviewers who participate in the discussion is the same for both scenarios, but the \emph{order in which reviewers join the discussion is different depending on the group the paper is allocated to}. Importantly, we underscore that these requirements do not imply that the strategies do not impact the set of reviewers who participate in the discussion of the paper. Instead, we only require that the set of participating reviewers is the same for both strategies (it may be different from the set of reviewers who discuss the paper when no strategy is applied).

The high-level idea of the discussion-management strategies we employ in the experiment is to use the na\"ive approach of requesting the most positive and the most negative reviewers to initiate the discussion for the corresponding groups of papers, and then complementing it with a balancing part which attempts to equalize the set of participating reviewers across the groups. Let us now introduce this idea in more detail.

To describe the discussion-management strategies, let us consider any participating paper such that not all reviewers gave the same score to this paper in their initial reviews. Also let $\posrevconv$ (respectively, $\negrevconv$) be a reviewer with the most positive (respectively, negative) opinion about the paper according to the initial reviews. In other words, $\posrevconv$ and $\negrevconv$ are reviewers who gave the highest and the lowest overall scores to the paper in the initial review. If more than one reviewer gave the highest (respectively, lowest) score, then $\posrevconv$ (respectively, $\negrevconv$) is selected by breaking ties uniformly at random.  Consider now the following two strategies, termed $\positive$ and $\negative$, towards managing the discussion of this paper: 
\begin{itemize}
     \item[$\positive$:] First, ask reviewer $\posrevconv$ to \emph{initiate} the discussion, then ask reviewer $\negrevconv$ to \emph{contribute} to the discussion
    \item[$\negative$:]  First, ask reviewer $\negrevconv$ to \emph{initiate} the discussion, then ask reviewer $\posrevconv$ to \emph{contribute} to the discussion
\end{itemize}

Both strategies consist of two parts: in the first part, which we call initiating, a reviewer with an extreme opinion about the paper is asked to initiate the discussion. In the second part, which we refer to as balancing, the reviewer with the score at another extreme is asked to contribute to the discussion. The strategies are similar in that they attempt to engage both the most positive and the most negative reviewers in the discussion of the paper. However, the crucial difference between $\positive$ and $\negative$ is the order in which reviewers are supposed to contribute: $\positive$ encourages the most positive reviewer to start the discussion while $\negative$ encourages the reviewer with the most negative opinion to initiate it. Importantly, both strategies proceed to requesting the second reviewer (after some waiting time) even if the first reviewer fails to fulfil the request.  

Intuitively, if we apply strategy $\positive$ to papers from one group and strategy $\negative$ to papers from the other, then we expect to observe a difference in opinions of initiating reviewers  between the groups, because in the initiating part of the strategies we target different kind of reviewers. Of course, the initiating part may simultaneously induce a difference in the populations of reviewers participating in the discussion. However, the balancing part of the strategies aims to compensate for it. Overall, by applying different strategies ($\positive$ versus $\negative$) to the two groups of papers, we expect to obtain a difference in the opinions of discussion initiators, keeping the population of reviewers participating in discussions the same in both groups. 

As a result, the intervention we perform in this experiment reduces the test for herding to a test for equality of acceptance rates between the two groups of papers. Indeed, if the herding effect is absent, the choice of strategy should not impact the outcome of papers disproportionately across the two groups treated with different strategies and hence we should not expect to observe a difference in acceptance rates between conditions. On the other hand, our intervention is designed to induce the largest possible difference between opinions of discussion initiators across the two groups. Given that under the presence of herding we expect the consensus opinion of reviewers to shift towards the opinion of the initiator of discussion, we expect the papers treated with strategy $\positive$ to enjoy a higher acceptance rate than their counterparts treated with $\negative$.

\smallskip

Of course, the reduction of the complex research question to a standard test of the difference in acceptance rates depends on the assumptions that mirror the idiosyncrasies formulated in Section~\ref{section:preliminaries}:

\begin{enumerate}[label=A\arabic*]
    \item \label{assumption:one} Our intervention does not introduce any difference across two groups of papers other than in the opinion of the discussion initiator (e.g., does not introduce a difference in the distributions of the participating reviewers or in some other characteristics of the discussion). 
    
    \item \label{assumption:two} Our intervention is successful in changing the order in which reviewers join the discussion of papers from different groups.
\end{enumerate} 
Intuitively, we designed the intervention to ensure that these assumptions are satisfied. However, a priori we cannot guarantee that these assumptions always hold. For example, Assumption~\ref{assumption:one} breaks if the balancing part of the strategies fails to equalize the populations of participating reviewers across conditions. In Section~\ref{section:analysis} we provide data-based evidence that support these assumptions.

\subsection{Details of the Experimental Procedure}
\label{sec:exppr}

Having designed the intervention, we now discuss how the set of participating papers was constructed and outline the details of the organization of the experiment.

\paragraph{Participating Papers.} To have a high detection power, we would like to run the experiment using all the papers submitted to the \ICMLyear{} conference. However, the issue with using all the papers is that some reviewers may be the most positive or the most negative reviewers for multiple papers, being overburdened with requests to initiate (contribute to) the discussion of these papers.

To limit the additional load on reviewers induced by our experiment, for each reviewer we limit the number of papers the reviewer is asked to initiate the discussion or contribute to the discussion to one each. This condition puts the limit on the number of papers we can use in the experiment. Consequently, to compensate for the potential decrease of power, we focus the scope of the experiment on the \emph{borderline} papers with some disagreement between reviewers as we expect the effect (if any) to be the most prominent in these papers. In particular, we do the following:
\begin{itemize}[itemsep=3pt, leftmargin=15pt, topsep=5pt]
    \item First, we use data from the past editions of the \ICML{} conference and scores given in initial reviews to identify the borderline papers: papers that are not clear accepts or clear rejects (more details on the choice of papers are given in Appendix~\ref{appendix:border}).
    
    \item Second, we exclude all papers that received identical overall scores in initial reviews from all reviewers as for these papers the notions of the most positive and most negative reviewers are undefined.
    
    \item Finally, from the papers not eliminated in the previous two steps, we find a subset of 1,544 participating papers $\exppap$ such that each reviewer is the most positive reviewer for at most one of these papers and the most negative reviewer for at most one of these papers.
\end{itemize}

For each strategy $\strategy \in \{\positive, \negative \}$, we refer to the reviewer who is asked to initiate the discussion as an \frst{} reviewer and the reviewer who is requested in the balancing part of the strategy as a \scnd{} reviewer. We now uniformly at random allocate papers from the set $\exppap$ into two groups subject to the aforementioned constraint that each reviewer is the \frst{} reviewer for at most one paper and the \scnd{} reviewer for at most one paper:
\begin{itemize}[leftmargin=35pt]
    \item[$\posgroup$:] The positive group which includes 755 papers treated with strategy $\positive$
    \item[$\neggroup$:] The negative group which includes 789 papers treated with strategy $\negative$
\end{itemize}
Having described the set of participating papers, we are now ready to discuss the timeline of the experiment.

\paragraph{Organization of the Experiment.}

\begin{figure}[t]
    \centering
    \includegraphics[width=14cm]{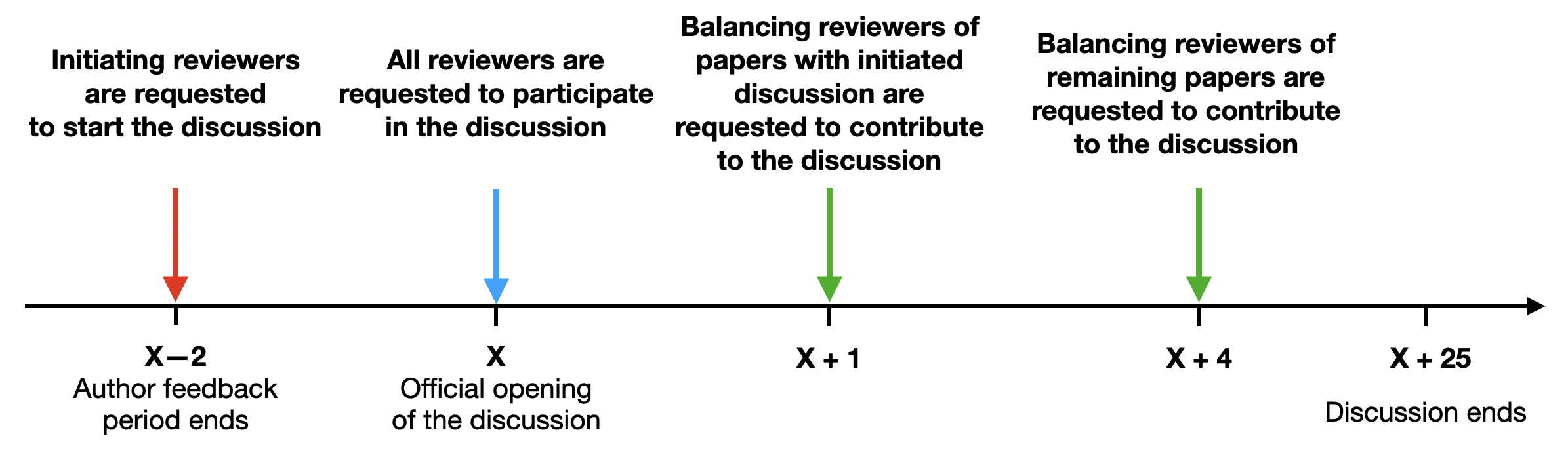}
    \caption{Timeline of the experiment. Day X is the day of the official discussion opening.}
    \label{fig:schedule}
\end{figure}

Figure~\ref{fig:schedule} depicts the pipeline implemented in the experiment. To ensure that \frst{} reviewers have sufficient time to initiate the discussion, we first open the discussion interface without notifying the general pool of reviewers, and send requests to \frst{} reviewers only. Two days after the initial requests, we notify all other reviewers that the discussion stage is open. Finally, we email \scnd{} reviewers in two stages: first, we send emails only for papers with initiated discussion, giving \frst{} reviewers of papers with no discussion initiated a little more time to start it. We then complete the intervention by targeting the remaining \scnd{} reviewers irrespective of whether the \frst{} reviewers of the corresponding papers fulfilled our request or not.

Through the first ten days of the experiment, we send reminders to the \frst{} and \scnd{} reviewers who have not fulfilled our request to initiate or contribute to the discussion of the corresponding papers. In order to avoid a disproportional impact on the discussion participants across two groups of papers, we ensure that the total number of reminders is the same for the \frst{} and \scnd{} reviewers. Importantly, the research hypothesis we evaluate in this paper may be sensitive to the awareness of the subjects, so in this study we employ deception and do not notify subjects (reviewers and area chairs) about the experiment.

Before we proceed to the results of the experiment, we note that three members of the study team were involved in the \ICML{} decision-making process. NS served as an area chair and AS and HD were program chairs. To avoid the conflict of interests, NS, AS and HD were not aware of what papers were used in the experiment. Moreover, we excluded papers chaired by NS from the analysis.

\section{Results of the Experiment}
\label{section:analysis}

In this section we present the results of the experiment. First, we report data that supports the assumptions we made in Section~\ref{sec:instrument}. In that, we begin with some general statistics on the discussion process that support Assumption~\ref{assumption:one} (Section~\ref{section:analysis:prelim}). We then discuss the efficacy of the intervention we employed (Section~\ref{section:analysis:eff}) and show that Assumption~\ref{assumption:two} is well-satisfied. Finally, we conclude with the analysis of the research question we study in this work (Section~\ref{section:analysis:fin}). 

\subsection{Preliminary Analysis (Data in Support of Assumption~\ref{assumption:one})}
\label{section:analysis:prelim}

Table~\ref{table:demography} provides some comparative statistics on the discussion process for the papers involved in the experiment and treated with strategies $\positive$ or $\negative$.  Observe in Table~\ref{table:demography} that the parameters of the discussion are similar across the two conditions (that is, similar for $\posgroup$ and $\neggroup$). This provides quantitative evidence that the randomization of papers to conditions occurred successfully and Assumption~\ref{assumption:one} is satisfied.  Rows 4 and 5 compare mean overall scores (the overall score takes integer values from 1 to 6 where larger values indicate higher quality) given by reviewers in the initial reviews, that is, before reviewers got to see the other reviews and the author feedback. Mean initial scores given by reviewers who participate in the discussion (Row 5) appear to be lower than mean scores computed over all reviewers (Row 4), suggesting that those who give lower scores are more active in discussing papers. However, there is no significant difference between the positive ($\posgroup$) and negative ($\neggroup$) groups of papers in these values. Hence, the data indicates that this trend is independent of the choice of the strategy. 

\begin{table}[htbp]
\begin{center}
{\sc Comparative Statistics on the Discussion Process}
\vskip 0.15in
\begin{small}
\begin{sc}
\begin{tabular}{lcr}
\toprule
          & $\posgroup$ & $\neggroup$ \\
\midrule
1. Number of papers 	                       	& 755	& 789	   \\ 
2. Fraction of papers with active discussion    & 0.97	& 0.97	   \\ 
3. Mean discussion length (\# messages)        & 4.41 	& 4.24 \\ 
4. Mean initial score (all revs)	            & 3.52 	& 3.52	   \\ 
5. Mean initial score (revs in discussion) 	& 3.44 	& 3.46     \\ 
6. Standard deviation of initial scores (all revs) 		    & 1.12 	& 1.11     \\ 
7. Fraction of papers with $\posrevconv$ active in discussion   & 0.79 & 0.79 \\ 
8. Fraction of papers with $\negrevconv$ active in discussion 	& 0.87 & 0.84 \\ 
9. Mean number of discussion participants (revs + area chairs) 	 	& 3.14 	& 3.06  \\ 
\bottomrule
\end{tabular}
\end{sc}
\end{small}
\end{center}
\vskip -0.1in
\caption{Comparison of some discussion statistics between papers treated with different discussion-management strategies. Except Row 5, all values are computed using all papers including those with no discussion. Permutation test at the level 0.05 (before multiple-testing adjustment) does not reveal significant differences between conditions.} 
\label{table:demography}
\end{table}

\begin{figure}[htbp]
    \centering
    \includegraphics[width=8cm]{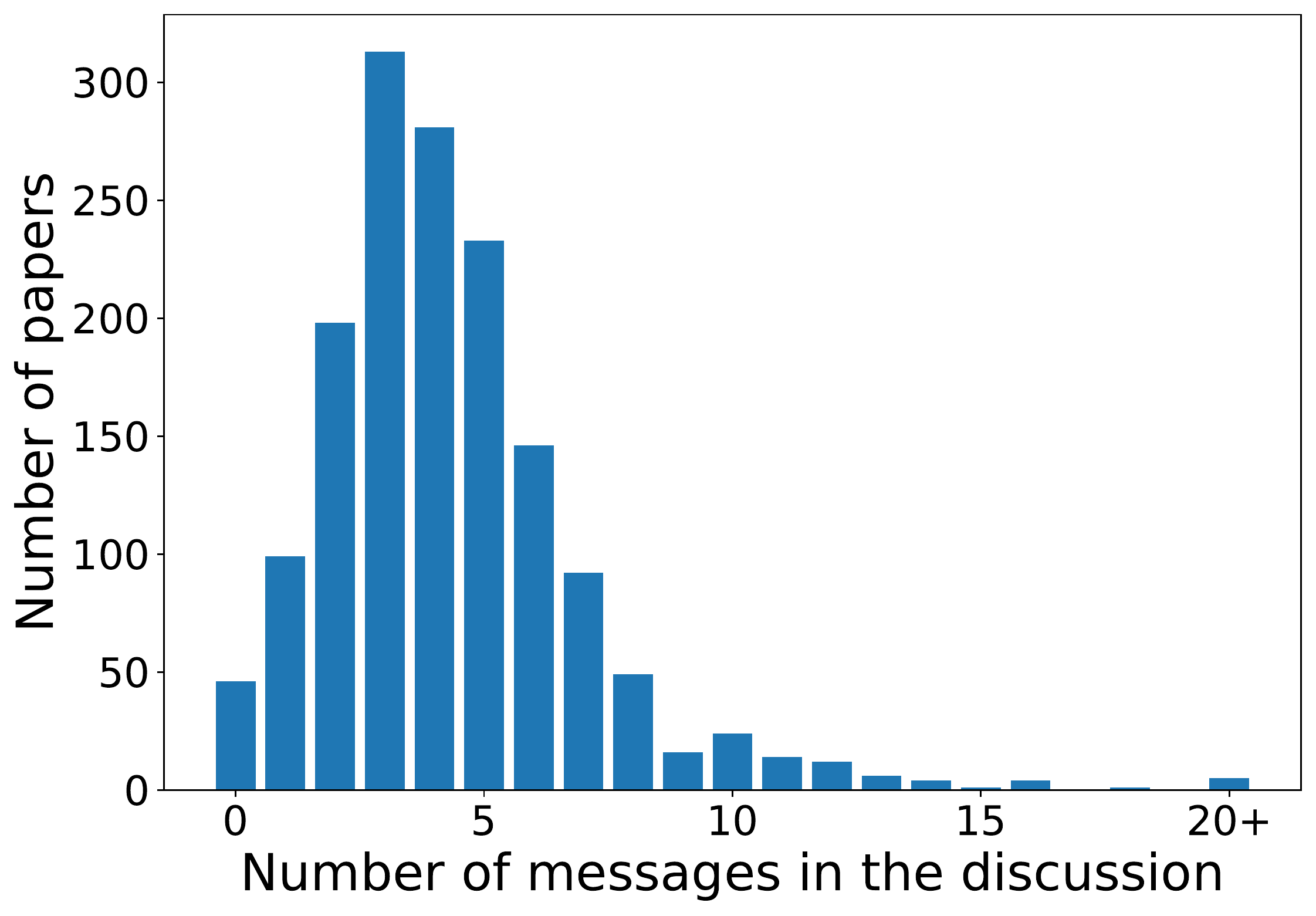}
    \caption{Distribution of the discussion length over all papers $\exppap$ used in the experiment.}
    \label{fig:disclen}
\end{figure}

Importantly, the activity of reviewers $\posrevconv$ and $\negrevconv$ in the discussion (Rows 7 and 8) is similar across the two groups of papers. This evidence shows that the choice of the discussion-management strategy does not introduce a difference across conditions in the distributions of reviewers who participate in the discussion. Finally, we observe that most of the papers used in the experiment had some discussion (Rows 2 and 3). Figure~\ref{fig:disclen} presents the distribution of the discussion length over all participating papers, revealing that a non-trivial amount of discussion takes place behind the scenes, thereby confirming that the experiment creates appropriate conditions for the herding behaviour to manifest.


\subsection{Efficacy of the Instrument (Data in Support of Assumption~\ref{assumption:two})}
\label{section:analysis:eff}

In the previous section we demonstrated that our intervention did not introduce a difference across conditions in metrics such as intensity of discussions and the population of participating reviewers. This observation supports Assumption~\ref{assumption:one} and indicates the appropriateness of our intervention.  However, in order for the experiment to be successful from the testing standpoint, the intervention needs to satisfy Assumption~\ref{assumption:two} and introduce a difference across conditions in the order in which reviewers join the discussion of the papers. Indeed, if all the emails we sent to reviewers were ignored (i.e., our attempt to impact the order failed), the subsequent analysis will not detect the phenomena even when the phenomena is present.

\begin{table}[b]
\begin{center}
{\sc Does the Intervention Affect who Initiates the Discussion?}
\vskip 0.15in
\begin{small}
\begin{sc}
\begin{tabular}{lccccc}
\toprule
          & $\posgroup$ & $\neggroup$ & $\diff$  & $\diff$ 95\% CI & $p$ value   \\
\midrule
1. Mean initial score (initiator)  & $4.03$ & $2.76$ & $1.27$ & $[1.15, 1.39]$ &  ${<.001}$    \\
2. Fraction of discussions initiated by $\posrevconv$ & $0.53$ 	& $0.09$ & $0.44$ & $[0.39, 0.48]$	& ${<.001}$	 \\
3. Fraction of discussions initiated by $\negrevconv$  & $0.15$ 	& $0.59$ & $-0.44$ & $[-0.48, -0.39]$ & ${<.001}$   \\
\bottomrule
\end{tabular}
\end{sc}
\end{small}
\end{center}
\vskip -0.1in
\caption{The impact of the intervention on who initiates the discussion. To compute values for Row 1, we use 1,140 papers for which (i) the discussion was initiated, and (ii) the discussion initiator was a reviewer (and not the area chair). For the last two rows, we use all papers including those with no discussion.  Bootstrapped confidence intervals are constructed for the difference of the relevant quantities between conditions. All $p$ values are computed using the permutation test with 10,000 iterations.} 
\label{table:efficiency}
\end{table}

Table~\ref{table:efficiency} reports relevant statistics and indicates a large difference between positive and negative groups of papers, suggesting that our intervention did indeed impact the order in which reviewers joined the discussion. Importantly, Row 1 demonstrates that initiators of discussions for papers from different groups have considerably different opinions about the papers. Overall, we conclude that our intervention is efficacious and satisfies the prerequisite for powerful testing (Assumption~\ref{assumption:two}).


\subsection{Main Analysis}
\label{section:analysis:fin}

Having demonstrated that the intervention we implemented in the experiment reasonably satisfies the required assumptions, we now continue to the analysis directly related to the research question we study in this work. Specifically, as we explained in the introduction and in Section~\ref{sec:design}, if herding behaviour exists, we expect it to manifest in the final decisions being disproportionately influenced by the opinion of the discussion initiator. Hence, given that for the positive group of papers $\posgroup$ the initial opinion of the discussion initiator was on average significantly more positive than that of initiators of discussions for the negative group of papers $\neggroup$, we expect to observe a disparity in the eventual acceptance rates between conditions.

Table~\ref{table:results} formalizes the intuition and performs the comparison of acceptance rates across papers treated with different strategies (Row 1). Additionally, Table~\ref{table:results} displays the updates of the scores made by reviewers (Rows 2--5).  First, the data does not indicate a statistically significant difference between acceptance rates in the two groups of papers ($\posgroup$ versus $\neggroup$).  Second, the data on the score updates suggests that in their final evaluations, reviewers tend to converge to the mean of initial independent opinions irrespective of the discussion-management strategy employed. Indeed, Row 2 demonstrates that the initiators of discussions update the scores towards the mean of all initial scores. Next, Rows 3 and 4 show that a significant update made by the discussion initiators is compensated by the update made by other reviewers, such that the overall amount of change in the mean scores is negligible. As expected, the outlined dynamics result in a significant decrease in the variance of scores per paper, but the effect is the same for both groups of papers (Row 5). 

Overall, the comparison of acceptance rates and changes in the mean reviewers' scores does not reveal any significant difference between the papers treated with different discussion-management strategies. Hence, we find no evidence of herding in the discussion phase of \ICMLyear{} peer review.

\begin{table}[t]
\begin{center}
{\sc Does the Intervention Affect the Outcome of Papers?}
\vskip 0.15in
\begin{small}
\begin{sc}
\begin{tabular}{lccccc}
\toprule
          & $\posgroup$ & $\neggroup$ & $\diff$ &  $\diff$ 95\% CI  & $p$ value  \\
\midrule
1. Acceptance rate & $0.21$	& $0.25$ & $-0.04$ & $[-0.08, 0.01]$	& $.122$    \\
2. Change in mean score (initiator)    & $-0.10$	& $0.20$ & $-0.30$ & $[-0.37, -0.23]$ & $<.001$\\
3. Change in mean  score (all revs)   & $0.01$	& $0.01$ & $0.00$ & $[-0.03, 0.04]$ & $.949$    \\ 
4. Change in mean  score (revs in discussion)   & $0.03$	& $0.02$ & $0.01$ & $[-0.04, 0.06]$ & $.697$    \\ 
5. Change in standard deviation of scores (all revs)  & $-0.23$ & $-0.21$ & $-0.02$ & $[-0.05, 0.02]$ & $.296$   \\ 

\bottomrule
\end{tabular}
\end{sc}
\end{small}
\end{center}
\vskip -0.1in
\caption{The impact of the intervention on the final outcome of papers. For Row 2, we use 1,140 papers for which (i) the discussion was initiated, and (ii) the discussion initiator was a reviewer (and not the area chair). For Row 4, we use papers with discussion. For all other rows, we use all papers including those with no discussion.   Bootstrapped confidence intervals are constructed for the difference of the relevant quantities between conditions. All $p$ values are computed using the permutation test with 10,000 iterations.} 
\label{table:results}
\end{table}

\section{Discussion}
\label{section:discussion}

The experiment we conducted in the present work aims at identifying the herding behaviour in the discussions of the \ICMLyear{} conference. The results presented in Section~\ref{section:analysis} show that while we managed to achieve an imbalance in the opinion of the discussion initiators across conditions, the difference in the acceptance rates is not significant and hence there is no evidence of herding. In this section, we provide an additional discussion on certain aspects of our experiment.

\subsection{Caveats Regarding the Design and Analysis of the Experiment}
\label{section:discussion:p1}

Given that our intervention induced the strong difference in the order in which reviewers joined the discussions (see Table~\ref{table:efficiency}), the absence of difference in  score updates (see Table~\ref{table:results}) allows us to conclude with a high degree of confidence that the choice of the discussion-management strategy ($\positive$ versus $\negative$) does not impact the way reviewers update their scores. Of course, herding (if present) does not necessarily need to manifest in how reviewers change their scores after the discussion. Instead, it can change some other characteristics such as what reviewers write in the textual messages which are later analyzed by the area and program chairs who make the final decisions. To account for these potential manifestations, we compared acceptance rates between the groups of papers (see Row 1 of Table~\ref{table:results}) and observed some difference in this quantity. However, this difference does not appear significant despite the large sample size we had in the experiment, suggesting that even if present, the effect has at most small size. That being said, we urge the reader to be aware of the following caveats.

\emph{Caveat 1. The design of the intervention.} Recall that our research question defines herding as a conditional dependence of the outcome of a paper on the choice of the discussion initiator. The test and the intervention we designed attempt to compare the outcomes of papers when the discussion is initiated by the most positive versus the most negative reviewers with the motivation that this difference is expected to be the largest in the presence of herding. Strictly speaking, the absence of a difference between these choices of the initiators does not imply the absence of the difference between any other choices of the initiators: for example, it is possible that the outcome of a paper would be impacted differently if we asked the reviewer with a non-extreme score to initiate the discussion. 

\emph{Caveat 2. The choice of papers.} As noted in Section~\ref{sec:exppr}, in this experiment we tried to identify a set of \emph{borderline} papers as these papers are more susceptible to the impact of the herding effect if it is present. However, our choice of the borderline papers was based on some indirect indicators and hence we could potentially fail to uncover the set of true borderline papers which would reduce the power of our test. 

To evaluate our choice of borderline papers, we use a rough classification of submissions into clear and borderline cases made by the area chairs. Note that this classification was performed after the discussion stage which could resolve the uncertainty present before the discussion stage when we selected the participating papers. Hence, the fraction of borderline papers in the area chairs' classification is a conservative estimate of the pre-discussion fraction of borderline papers. Nonetheless, 30\% of submissions used in the experiment were classified by the area chairs as borderline cases in contrast to 18\% of those not involved in the experiment ($\Delta = 0.12, p = .002$). Hence, our choice of the borderline papers was better than random and the set of the participating papers $\exppap$ contained a large fraction of papers for which the decisions were not clear before the discussion.

\emph{Caveat 3. Validity of Assumption~\ref{assumption:one}.} The validity of the conclusions we make is based on the assumptions formulated in Section~\ref{sec:instrument}. Note that a violation of the assumptions could increase the false alarm probability or could reduce the power of the test. The data we presented in Section~\ref{section:analysis:prelim} and Section~\ref{section:analysis:eff} strongly supports Assumptions~\ref{assumption:one} and~\ref{assumption:two}. However, as a note of caution, we remark that there is some space for potential violations of Assumption~\ref{assumption:one}. Indeed, in Table~\ref{table:demography} we establish that the marginal values of relevant indicators of discussion activity are similar across groups. However, this observation does not imply that the value of these indicators for each paper would not change if that paper was placed in the other condition (recall the thought experiment in Section~\ref{sec:instrument}). Hence, the the outcome of this study should be considered together with this opportunity for the violation of Assumption~\ref{assumption:one}.

\emph{Caveat 4. Spurious correlations induced by reviewer identity.} In peer review, each reviewer participates in the discussion of multiple papers. Similarly, each area chair manages several papers. Hence, strictly speaking, the outcomes of two papers that have at least one reviewer in common  (are managed by the same area chair) may not be statistically independent due to correlations introduced by the reviewer (area chair) identities. This issue puts a strain on the testing procedure because in contrast to the vanilla A/B testing framework which assumes that samples are independent of each other, in our case we receive correlated samples. In the domain of empirical studies of the peer-review procedure~\citep{shah2017design, tomkins17wsdm, lawrence14neurips} such spurious correlations are usually tolerated, because otherwise the sample size would be negligible. Additionally, simulations performed by~\citet{stelmakh2019testing} demonstrate that unless reviewers are involved in the discussion of dozens of submissions, the impact of such spurious correlations is limited. 

Nevertheless, in this work we take some additional steps to minimize the impact of these spurious correlations. To this end, we simultaneously also perform the analysis on a subset of 937 papers $\specexppap \subset \exppap$ which is constructed such that each reviewer is the the most positive or the most negative reviewer for at most one paper from set $\specexppap$ \emph{in total}.\footnote{Compare to $\exppap$ which is constructed such that each reviewer is the the most positive or the most negative reviewer for at most one paper \emph{each}.} This additional reduction of the sample size allows us to limit the impact of the reviewer identity on the outcome of submissions. Of course, by doing so we do not guarantee that there is no reviewer who participates in the discussion of more than one paper from the set $\specexppap$, but we guarantee that the discussion participants who are targeted by our strategies are unique.  Appendix~\ref{appendix:border} gives more details on how sets $\exppap$ and $\specexppap$ were constructed. The results of this additional analysis are presented in Appendix~\ref{appendix:moreeval} and lead to the same conclusions as in Section~\ref{section:analysis}.

\emph{Caveat 5. Opinion of the discussion initiator.}  In this work we used the scores given by reviewers in the initial reviews to infer the pre-discussion opinion of reviewers and assumed that reviewers begin the discussion from advocating these opinions. However, the fact that a reviewer has some pre-discussion opinion does not guarantee that they advocate the same position in the discussion because the latter is also influenced by other reviewers' reviews and the author feedback.  Indeed, past research~\citep{teplitskiy19influence} suggests that reviewers do listen to each other and may update their initial independent opinions in light of opinions expressed in initial reviews of other reviewers. The data we obtained in the experiment suggests that while such updates take place, their magnitude is small enough and does not break our intervention. Indeed, Row 2 of Table~\ref{table:results} and Row 1 of Table~\ref{table:efficiency} indicate that reviewers with extreme pre-discussion opinions remain on the different sides of the mean pre-discussion group opinion even according to the final scores. Thus, we conclude that our intervention succeeded in creating a difference in opinions of discussion initiators across groups.

\emph{Caveat 6. Alternative model of herding.} In this paper we assume that the herding behaviour in peer review manifests in final decisions being moved towards the position of the reviewer who initiates the discussion. However, the data presented in Table~\ref{table:results} shows that initiators of the discussion tend to slightly update their scores towards the mean of initial scores given by all reviewers. Hence, an alternative model of the herding behaviour is that the sentiment demonstrated by the initiating reviewer carries over to other reviewers who could change their behaviour accordingly. For example, the positive score update of initiators with a negative initial opinion may demonstrate a positive sentiment, which could affect opinions of other reviewers in a positive way. Under this alternative model of herding, we would expect papers from the negative group $\neggroup$ to enjoy a higher acceptance rate than their counterparts from the positive group $\posgroup$. While this agrees with the observed acceptance rates reported in Table~\ref{table:results}, we reiterate that the difference between the acceptance rates is not significant and the effect size is small, so our test does not provide evidence in support of this alternative model either.

\subsection{Relation to Past Work}

Past literature suggests the presence of herding behaviour in group discussions. \citet{McGuire87group} observe that the first solution proposed to a group predicts the group decision better than an aggregate of initial opinions independently expressed in a pre-discussion survey. \citet{dubrovsky91status} document an impact of the interplay between the status of discussion participants and the opinion of the group member who proposed the first concrete solution on the final group decision. Closest to the present work, \citet{weisband92advocacy} further investigates the herding effect in a semi-randomized controlled trial and declares that the initiators of discussion manage to influence the group opinion when they step in after an initial general discussion of the problem  that is, when they have some understanding of the general opinions of other discussants, but no concrete decisions have been proposed.

In contrast, in this experiment, we did not detect herding in the peer-review discussion. Let us now discuss the relationship of the current experiment to these past works. First, we note that the papers of~\citet{McGuire87group} and~\citet{dubrovsky91status} study  the herding effect when the discussion initiators are self-selected. The difference between the self-selected and assigned initiators appears to be significant, because the former may be associated with other personal qualities such as assertiveness and energy. Hence, our work is not directly comparable to these studies as we attempt to randomize the identity of the discussion initiator.

The experiment of~\citet{weisband92advocacy} employs randomization and in addition to the self-selection scenario considers the setup in which the first person to propose the solution to the group is chosen uniformly at random. This work finds that the randomly assigned initiator exerts much smaller influence on the group decision than the self-selected initiator. We caveat however, that in the experiment of~\citet{weisband92advocacy}, the fact that the initiator is selected at random was known to the whole group before the beginning of the discussion. Hence, it is plausible that other group members did not perceive the initiator as a leader and could adjust their behaviour accordingly~\citep{weisband92advocacy}. In contrast, in the present experiment the non-initiating reviewers were not aware of the intervention and hence from their point of view the assigned initiator of the discussion possessed all the properties of the self-selected initiator~\citep{hollander1978leadership}.

Finally, there is a subtle difference between the definition of the herding effect made by~\citet{weisband92advocacy} and the definition we use in this paper. According to~\citet{weisband92advocacy}, the herding is present when the first solution formulated in the group discussion predicts the group final decision better than the mean of the pre-discussion independent opinions. Note that according to this definition, the herding may be present even if the first solution proposed to the group is independent of who is selected to formulate this opinion, that is, even when all discussants would propose the same solution should they be selected to start the discussion. In contrast, in our settings it is natural to define herding to be present only when the opinion of the discussion initiator is different depending on who is selected to initiate the discussion, because the goal of the present work is to inform the area chairs about the potential consequences of their discussion initiating strategy.

In addition to the aforementioned distinctions from the past work, we note that in the present experiment reviewers are engaged in a much more analytical task as compared to the previous works in which some toy problems were used to study the discussion dynamics. Hence, the absence of the herding behaviour in peer review may be due to the fact that reviewers have a rational mindset which is hypothesized to reduce a reliance on heuristics responsible for various cognitive biases~\citep{stanovich99rational, kahneman02representativeness}.

Beyond testing for herding, in this paper we also document effects predicted by past works on discussion in peer review~\citep{teplitskiy19influence, hofer00care}: reviewers tend to update their scores towards the consensus pre-discussion opinion, and the discussion increases the agreement among reviewers. Coupled with the observation that an increased agreement does not necessarily result in an increased accuracy of the decision~\citep{hofer00care}, our observations highlight an importance of additional research on the discussion dynamics in peer review.


\section*{Acknowledgments}

We thank Edward Kennedy for  useful comments on the initial design of this study. We are also grateful to the support team of the Microsoft Conference Management Toolkit (CMT) for their continuous support and help with multiple customization requests. Finally, we appreciate the efforts of all reviewers and area chairs involved in the \ICMLyear{} review process. This study was approved by Carnegie Mellon University Institutional Review Board.

This work was supported in part by NSF CAREER award 1942124 and in part by NSF CIF 1763734.

\bibliography{references.bib}
\bibliographystyle{apalike}

\newpage

\appendix

\noindent \textbf{\Large{Appendix}}

\medskip

\noindent We provide supplementary materials and additional discussion. In Appendix~\ref{appendix:border} we formalize the rules we used to identify the set of borderline papers when constructing the set of participating papers $\exppap$ and $\specexppap$ introduced in Section~\ref{sec:exppr} and Section~\ref{section:discussion:p1}, respectively.  Appendix~\ref{appendix:moreeval} is dedicated to the additional analysis on the subset of papers $\specexppap$.


\section{Construction of Sets $\exppap$ and $\specexppap$}
\label{appendix:border}

In this section we specify how the sets $\exppap$ (introduced in Section~\ref{sec:exppr}) and $\specexppap$ (introduced in Caveat 4, Section~\ref{section:discussion:p1}) of participating papers were constructed from the set of all $\numpap = 4625$ papers not withdrawn from \ICMLyear{} by the beginning of the discussion period.

We begin from set $\exppap$ and recall that our goal is to identify a subset of borderline papers such that each reviewer is the most positive reviewer for at most one paper from this subset and the most negative reviewer for at most one paper from this subset. To this end, we perform the following two-step procedure:

\paragraph{Step 1.} First, we identify the set of borderline papers as follows. The overall scores given in the initial reviews were in the set $\{1, 2, \ldots, 6\}$ so for each paper $\papidx \in [\numpap]$, we let $\papload_{\papidx}$ to denote the number of reviewers assigned to the paper (typically $\papload_{\papidx}$ equals 3 or 4) and let $(\score_1, \score_2, \ldots, \score_{\papload_{\papidx}}) \in \{1, 2, \ldots, 6\}^{\papload_{\papidx}}$ to denote the collection of overall scores given to paper $\papidx$ in initial reviews.\footnote{Here, we adopt the standard notation $[\nu] = \left\{1, 2, \ldots, \nu \right\}$ for any positive integer $\nu$.}  With this notation, using acceptance statistics of the \ICML{} 2019 conference, we construct a set of borderline papers $\tempset$ by identifying  submissions that satisfy the following criteria: 

    \begin{enumerate}[noitemsep, label=C\arabic*]
        \item \label{cr:1} The mean overall score is such that in \ICML{} 2019 the paper is in the borderline category:
        \begin{align*}
            \frac{1}{\papload_{\papidx}} \sum\limits_{\revidx = 1}^{\papload_{\papidx}} \score_{\revidx} \in [2.7, 4.5].
        \end{align*}

        \item \label{cr:2} The minimum and maximum overall scores are on the different sides of the decision spectrum: 
        \begin{align*}
            \max(\score_1, \score_2, \ldots, \score_{\papload_{\papidx}}) \ge 4 \quad \text{and} \quad \min(\score_1, \score_2, \ldots, \score_{\papload_{\papidx}}) \le 3.
        \end{align*}
    \end{enumerate}
    Note that for each borderline paper $\papidx \in \tempset$ we are guaranteed that there is some disagreement between reviewers.
    
\paragraph{Step 2.} Having the set of borderline papers $\tempset$ defined, we construct $\exppap$ by greedily finding a subset of $\tempset$ that satisfies the requirement of each reviewer being the most positive reviewer for at most one paper from this subset and the most negative reviewer for at most one paper from this subset.

\medskip

Let us now proceed to the definition of set $\specexppap \subset \exppap$ which has an additional property of each reviewer being the most positive or the most negative reviewer for at most one paper in total. This set is also constructed by greedily finding a subset of $\exppap$ that satisfies this requirement. However, to facilitate tie-breaking, we additionally request that for each paper $\papidx \in \specexppap$ the most positive and most negative reviewers disagree in the initial reviews by at least 2 points:
\begin{align*}
        \max(\score_1, \score_2, \ldots, \score_{\papload_{\papidx}}) - \min(\score_1, \score_2, \ldots, \score_{\papload_{\papidx}}) \ge 2.
\end{align*}
Hence, set $\specexppap$ has an additional property of containing papers with high disagreement between reviewers in the initial reviews.

\section{Additional Analysis}
\label{appendix:moreeval}

Recall that the set of papers $\specexppap$ (defined in Appendix~\ref{appendix:border}) constitutes a subset of all the participating papers $\exppap$ and additionally satisfies the constraints of (i) each reviewer serving as the \frst{} or \scnd{} reviewer for at most one paper from $\specexppap$ in total and (ii) each paper $\papidx \in \specexppap$ being such that the most positive and the most negative reviewers disagree in the initial reviews by at least two points. In this section, we replicate the analysis presented in Section~\ref{section:analysis} conditioning on this subset of papers.

\begin{table}[htbp]
\begin{center}
{\sc Comparative Statistics on the Discussion Process}
\vskip 0.15in
\begin{small}
\begin{sc}
\begin{tabular}{lcr}
\toprule
          & $\specposgroup$ & $\specneggroup$ \\
\midrule
1. Number of papers 	                       	& 460 	& 477	   \\ 
2. Fraction of papers with active discussion    & 0.96 	& 0.98	   \\ 
3. Mean discussion length (\# messages)         & 4.52 	& 4.25  \\ 
4. Mean initial score (all revs)	            & 3.51 	& 3.52	   \\ 
5. Mean initial score (revs in discussion) 	    & 3.42	& 3.45     \\ 
6. Standard deviation of initial scores (all revs) 		    & 1.22	& 1.20    \\ 
7. Fraction of papers with $\posrevconv$ active in discussion   & 0.78 & 0.79 \\ 
8. Fraction of papers with $\negrevconv$ active in discussion 	& 0.87 	& 0.85 \\ 
9. Mean number of discussion participants (revs + area chairs) 	& 3.16	& 3.10  \\ 
\bottomrule
\end{tabular}
\end{sc}
\end{small}
\end{center}
\vskip -0.1in
\caption{Comparison of some discussion statistics between papers treated with different discussion-management strategies, conditioned on the subset of papers $\specexppap$. Except Row 5, all values are computed using all papers including those with no discussion. Permutation test at the level 0.05 (before multiple-testing adjustment) does not reveal significant differences between conditions.} 
\label{table:demography2}
\end{table}

Mirroring the analysis on the full set of participating papers (Table~\ref{table:demography}), Table~\ref{table:demography2} indicates that the parameters of the discussion are similar across the two conditions. Hence, we also conclude that data supports Assumption~\ref{assumption:one} and the intervention did not result in a difference across conditions in the distributions of reviewers who participate in the discussion.

\begin{table}[b]
\begin{center}
{\sc Does the Intervention Affect who Initiates the Discussion?}
\vskip 0.15in
\begin{small}
\begin{sc}
\begin{tabular}{lccccc}
\toprule
          & $\specposgroup$ & $\specneggroup$ & $\diff$  & $\diff$ 95\% CI & $p$ value   \\
\midrule
1. Mean initial score (initiator)  & $4.09$ & $2.69$ & $1.40$ & $[1.24, 1.55]$ &  ${<.001}$   \\
2. Fraction of discussions initiated by $\posrevconv$ & $0.53$ 	& $0.11$ & $0.42$ & $[0.36, 0.47]$	& ${<.001}$	 \\
3. Fraction of discussions initiated by $\negrevconv$  & $0.16$ 	& $0.60$ & $-0.44$ & $[-0.49, -0.38]$ & ${<.001}$    \\
\bottomrule
\end{tabular}
\end{sc}
\end{small}
\end{center}
\vskip -0.1in
\caption{The impact of the intervention on who initiates the discussion, conditioned on the subset of papers $\specexppap$. To compute values for Row 1, we use 1,140 papers for which (i) the discussion was initiated, and (ii) the discussion initiator was a reviewer (and not the area chair). For the last two rows, we use all papers including those with no discussion.  Bootstrapped confidence intervals are constructed for the difference of the relevant quantities between conditions. All $p$ values are computed using the permutation test with 10,000 iterations.} 
\label{table:efficiency2}
\end{table}

Next, we investigate the efficacy of our intervention and proceed to Table~\ref{table:efficiency2} that compares relevant statistics. Observe that the values in Table~\ref{table:efficiency2} are very similar to those reported in Table~\ref{table:efficiency}, suggesting that the intervention continues to introduce the required difference in opinions of discussion initiators between the groups of papers even when we zoom in on the target subset of papers $\specexppap$. Thus, we conclude that Assumption~\ref{assumption:two} continues to hold.

Having confirmed the efficacy of the intervention, we proceed to the comparison of the outcomes of submissions. The results presented in Table~\ref{table:results2} mimic those reported in Table~\ref{table:results}. The most notable difference is that the significance of the difference in acceptance rates becomes closer to the threshold of $0.05$, but still does not cross it. Given that the number of submissions involved in the experiment is large, we conclude that we do not observe strong evidence of the herding behaviour even after conditioning on the set of papers $\specexppap$.

\begin{table}[t]
\begin{center}
{\sc Does the Intervention Affect the Outcome of Papers?}
\vskip 0.15in
\begin{small}
\begin{sc}
\begin{tabular}{lccccc}
\toprule
          & $\specposgroup$ & $\specneggroup$ & $\diff$ &  $\diff$ 95\% CI  & $p$ value  \\
\midrule
1. Acceptance rate & $0.21$	& $0.26$ & $-0.05$ &  $[-0.11, 0.00]$ & $.079$    \\
2. Change in mean score (initiator)    & $-0.11$ & $0.21$ & $-0.32$ & $[-0.42, -0.22]$ & $<.001$ \\
3. Change in mean  score (all revs)   & $0.01$ & $0.01$ & $0.00$ & $[-0.05, 0.05]$ & $.925$        \\ 
4. Change in mean  score (revs in discussion)   & $0.02$ & $0.02$ & $0.00$ & $[-0.06, 0.07]$ & $.867$   \\ 
5. Change in standard deviation of scores (all revs)  & $-0.26$ & $-0.25$ & $-0.01$ & $[-0.06, 0.03]$ & $.560$    \\ 

\bottomrule
\end{tabular}
\end{sc}
\end{small}
\end{center}
\vskip -0.1in
\caption{The impact of the intervention on the final outcome of papers, conditioned on the subset of papers $\specexppap$. For Row 2, we use 1,140 papers for which (i) the discussion was initiated, and (ii) the discussion initiator was a reviewer (and not the area chair). For Row 4, we use papers with discussion. For all other rows, we use all papers including those with no discussion.  Bootstrapped confidence intervals are constructed for the difference of the relevant quantities between conditions. All $p$ values are computed using the permutation test with 10,000 iterations.} 
\label{table:results2}
\end{table}

\end{document}